\pdfoutput=1
\documentclass[namedreferences]{solarphysics}
\usepackage[optionalrh]{spr-sola-addons} 
\usepackage{graphicx}        
\usepackage{color}           
\usepackage{url}             
\usepackage{subfig}

\newcommand{\aap}{    {\it Astron. Astrophys.}}

\newcommand{\apj}{    {\it Astrophys. J.}}
\newcommand{\apjl}{    {\it Astrophys. J. Lett.}}

\newcommand{\araa}{   {\it Ann. Rev. Astron. Astrophys.}}

\newcommand{\jgr}{    {\it J. Geophys. Res.}}

\newcommand{\solphys}{{\it Solar Phys.}}

\begin{document}

\begin{article}

\begin{opening}

%

  \title{How Many CMEs Have Flux Ropes? Deciphering the Signatures
	 of Shocks, Flux Ropes, and Prominences in Coronagraph
	 Observations of CMEs }

\author{A.~\surname{Vourlidas}$^{1}$\sep
	B.J.~\surname{Lynch}$^{2}$\sep  
        R.A.~\surname{Howard}$^{1}$\sep
	Y.~\surname{Li}$^{2}$      
       }
\runningauthor{Vourlidas et al.}
\runningtitle{Flux Rope CMEs}

\institute{
$^{1}$ Space Sciences Division, Naval Research Laboratory, Washington DC, USA \\
$^{2}$ Space Sciences Laboratory, University of California, Berkeley, CA, USA \\
              }

\begin{abstract}
  We intend to provide a comprehensive answer to the question on
  whether all Coronal Mass Ejections (CMEs) have flux rope
  structure. To achieve this, we present a synthesis of the LASCO CME
  observations over the last sixteen years, assisted by 3D MHD
  simulations of the breakout model, EUV and coronagraphic
  observations from \textsl{STEREO} and \textsl{SDO}, and statistics
  from a revised LASCO CME database. We argue that the bright loop
  often seen as the CME leading edge is the result of pileup at the
  boundary of the erupting flux rope irrespective of whether a cavity
  or, more generally, a 3-part CME can be identified. Based on our
  previous work on white light shock detection and supported by the
  MHD simulations, we identify a new type of morphology, the
  `two-front' morphology. It consists of a faint front followed by
  diffuse emission and the bright loop-like CME leading edge. We show
  that the faint front is caused by density compression at a wave (or
  possibly shock) front driven by the CME. We also present
  high-detailed multi-wavelength EUV observations that clarify the
  relative positioning of the prominence at the bottom of a coronal
  cavity with clear flux rope
  structure. 
  Finally, we visually check the full LASCO CME database for flux rope structures. In the
  process, we classify the events into two clear flux rope classes
  (`3-part', `Loop'), jets and outflows (no clear structure). We find
  that at least 40\% of the observed CMEs have clear flux rope
  structures and that $\sim29\%$ of the database entries are
  either misidentifications or inadequately measured and should be
  discarded from statistical analyses. We propose a new definition for
  flux rope CMEs (FR-CMEs) as a coherent magnetic, twist-carrying
  coronal structure with angular width of at least 40$^\circ$ and able
  to reach beyond 10 R$_{\odot}$ which erupts on a time scale of a few
  minutes to several hours We conclude that flux ropes are a common
  occurrence in CMEs and pose a challenge for future studies to
  identify CMEs that are clearly \textsl{not\/} FR-CMEs.

\end{abstract}
\end{opening}

\section{Introduction} \label{sec:Introduction}

Since their detection in the early 1970s, Coronal Mass Ejections (CMEs) have
been the subject of intense investigation with regard to their
initiation mechanisms, their effects on the corona and their
association with other coronal phenomena (eg., flares and
prominences). This \textsl{Topical Issue\/} presents results from a
Coordinated Data Analysis Workshop (CDAW) devoted to the question: `Do
All Coronal Mass Ejections (CMEs) Have Flux Rope Structures?' Such a
specific physics-based question shows that we have come a long way
towards understanding the nature of these explosive events especially
when we consider the original definition of a CME: `a relatively short
scale white light feature propagating in a coronagraph's field of view'
(paraphrasing \opencite{1984JGR....89.2639H}).

Traditionally, CMEs were observed with visible light coronagraphs and
clues on their origin and nature were based on their morphology in
those images
\cite{1979SoPh...61..201M,1985JGR....90.8173H,1993STIN...9326556B}. Despite
the apparently large variation in the appearance of CMEs, two
particular morphologies stand out: the `loop'-CME where a bright
narrow loop-like structure comprises the CME front, and the
`3-part'-CME \cite{1985JGR....90..275I} where the bright front is
followed by a darker cavity which frequently contains a bright
core. It has become the archetypical morphology of a CME even though
the `3-part' morphology could be identified in only about a third of
the events \cite{1979SoPh...61..201M}. It is still unclear whether the
remaining variation is the result of projection effects due to the
optically thin nature of the emission or not.

It was recognized early that the cavity rather than the prominence in
the core drove the CME \cite{1987sowi.conf..181H}. An initial controversy on whether CMEs were planar (i.e., ejected loops) or
three-dimensional (i.e., bubbles) structures was largely resolved by the end of the 1980's.
\inlinecite{1983SoPh...83..143C} demonstrated, using polarization
analysis, that the loop front was indeed a bubble. The identification
of halo CMEs by \inlinecite{1982ApJ...263L.101H} with their
quasi-circular appearance established their three-dimensional (3D)
nature and led to the adaption of the 'ice-cream' model to describe
and fit the kinematics of these events \cite{1982ApJ...263L.101H,2002JGRA..107.1223Z,2004JGRA..10903109X,2005JGRA..11008103X}. A
bubble or spherical structure is the intrinsic assumption behind this
model which, by the way, is not a proper description as we will discuss later.

As theories progressed towards a more physical basis for the CME
initiation, they focused on 3D magnetic topologies that could account
for the `3-part' morphology and the frequent association with
prominences. This quickly led to scenarios of rising loop arcades,
overlying a prominence, which underwent reconnection to form magnetic
flux ropes (FR, hereafter;
\opencite{1982SoPh...79..129A}; \opencite{1990JGR....9511919F}). Alternatively,
the FR could pre-exist and rise under the driving of Lorenz forces
\cite{1974A&A....31..189K,1993GeoRL..20.2319C}. While the question on
whether the FR is formed before or during the eruption remains open,
the overwhelming majority of magnetohydrodynamics (MHD) models and
simulations agree on one thing. Namely, the erupting structure is
always a FR \cite{2011LRSP....8....1C}. There is no physical mechanism that can produce
a large-scale fast eruption from the corona without ejecting a fluxrope, to the
best of our knowledge.

At the same time, in-situ measurements of interplanetary CMEs (ICMEs)
often encounter structures with smooth rotation in one, or more,
components of the magnetic field which can be fitted with FR models
(\opencite{1982JGR....87..613K}; \opencite{1990JGR....9511957L};
\opencite{2011SoPh..273..205I}, to name a few). These so-called
Magnetic Clouds (MCs) can be considered then as the interplanetary
manifestations of the ejected FR predicted by theory and possibly
detected as the cavity in the `3-part'-CMEs \cite{1982GeoRL...9.1317B}.
\inlinecite{2003JGRA..108.1156C} found that 100\% of ICMEs detected
during solar minimum were MCs reducing to $<20\%$ during solar maximum.

So the CDAW question regarding the nature of CMEs, at least in the
case of `3-part'-CMEs, seems to have been answered. A CME is 
simply the ejection of a magnetic FR structure from the lower corona which
takes the form of a `3-part'-CME or a MC depending on the instrumenation
used (images or in-situ, respectively) to detect it.

But, if a FR is a necessary ingredient for an ejection, why not all
CMEs show evidence for such structure? In other words, why all CMEs
are not `3-part'-CMEs? Some have just a loop front while others
appear as jets or structureless clouds or blobs. For example,
\inlinecite{1985JGR....90.8173H} categorized CMEs, between 6-10
R$_{\odot}$, into ten morphological classes based on their appearance
in \textsl{Solwind\/} observations. Why is there such a large variety
of shapes? Could there be other types of magnetic structures, besides
FRs, ejected from the Sun? If they do exist, they would suggest a
major gap in our understanding of eruptive processes, given the
prevalence of FR in our theories.

Second, not all ICMEs exhibit MC signatures. Is this simply a result
of `glancing' cuts between in-situ instruments and the ICME? Or do
CME FRs lose their coherence as they travel in the interplanetary
space, through reconnection with the ambient solar wind for example \cite{2007SoPh..244..115D}?

Third, many fast ICMEs are driving a shock followed by a
sheath of post-shocked plasma. The resulting five-part ICME (shock,
sheath, dense front, cavity, and dense plug) does not have a coronal
counterpart. Where are the five-part CMEs or more precisely, where are
the shock and sheath signatures in the coronagraph images? Shocks
could deflect streamers and generally affect the ambient corona, ahead
and at the flanks of a CME, thus creating complex brightness
distributions in the images.  Could such effects be responsible for
misidentifications, and hence misinterpretations, of CME morphologies,
kinematic profiles, and associations with structures in the low corona
or the inner heliosphere?

Fourth, and related point, the emission processes in both low (EUV) and middle (white
light) corona are optically thin resulting in images that are
projections on the plane of sky (POS). Do these projections affect our
ability to properly interpret observations and how can we account for
them? We will address this problem throughout this paper.

The Large Angle and Spectrometric Coronagraph (LASCO;
\opencite{1995SoPh..162..357B}) project has accumulated the largest and longest
database of coronagraphic observations of CMEs since 1996. Spanning more than 
a complete solar cycle, it is reasonable to expect that events of every
possible orientation, size, speed, mass, and morphologies have
been captured. We should be in position to understand the role of
projection effects on the images, identify the origin of
the various features (CME or not) in a given LASCO image, and hence
answer the question posed in this \textsl{Topical Issue}.

To accomplish this task comprehensively we have given this paper
a relatively large scope. It represents a synthesis of the observational
knowledge gained over the sixteen years of LASCO observations. In the
following sections, we will provide: evidence for the FR structure
within CME cavities (Section~\ref{sec:3part}), evidence for the
existence of white-light shock and tips on distinguishing the shock
front from the CME front (Section~\ref{sec:shock}), theoretical
support for these interpretations using synthetic images from
3D MHD simulations (Section~\ref{sec:models}),
observations that clarify the connection between prominence and
erupting cavity (Section~\ref{sec:prom}), and finally statistics on
the occurence of `3-part' or more precisely FR-CMEs, along with a
discussion on the constrains of event lists
(Section~\ref{sec:stats}). We discuss and conclude in
Section~\ref{sec:discussion}.

We will support several of our predictions and conclusions by using
two-viewpoint imaging afforded by the Sun-Earth Connection Coronal and
Heliospheric Investigation (SECCHI; \opencite{2008SSRv..136...67H})
on-board the {\it Solar TErrestial RElations Observatory (STEREO)}
\cite{2008SSRv..136....5K}. We will use the SECCHI observations as
necessary but we want to focus on the single viewpoint from LASCO for
two reasons. First, this article is part of a workshop devoted on the
analysis of events observed with LASCO. Second, and more important,
the \textsl{STEREO\/} mission has a finite lifetime. Budgetary and
other concerns suggest that future observations (whether research or
operationally oriented) will be obtained from a single vantage
point. It is therefore crucial that future observers can interpret
such single viewpoint observations accurately.

\section{Where is the Flux Rope? The 3-part CME} \label{sec:3part} 

\begin{figure}
\centerline{
\includegraphics[width=0.8\textwidth]{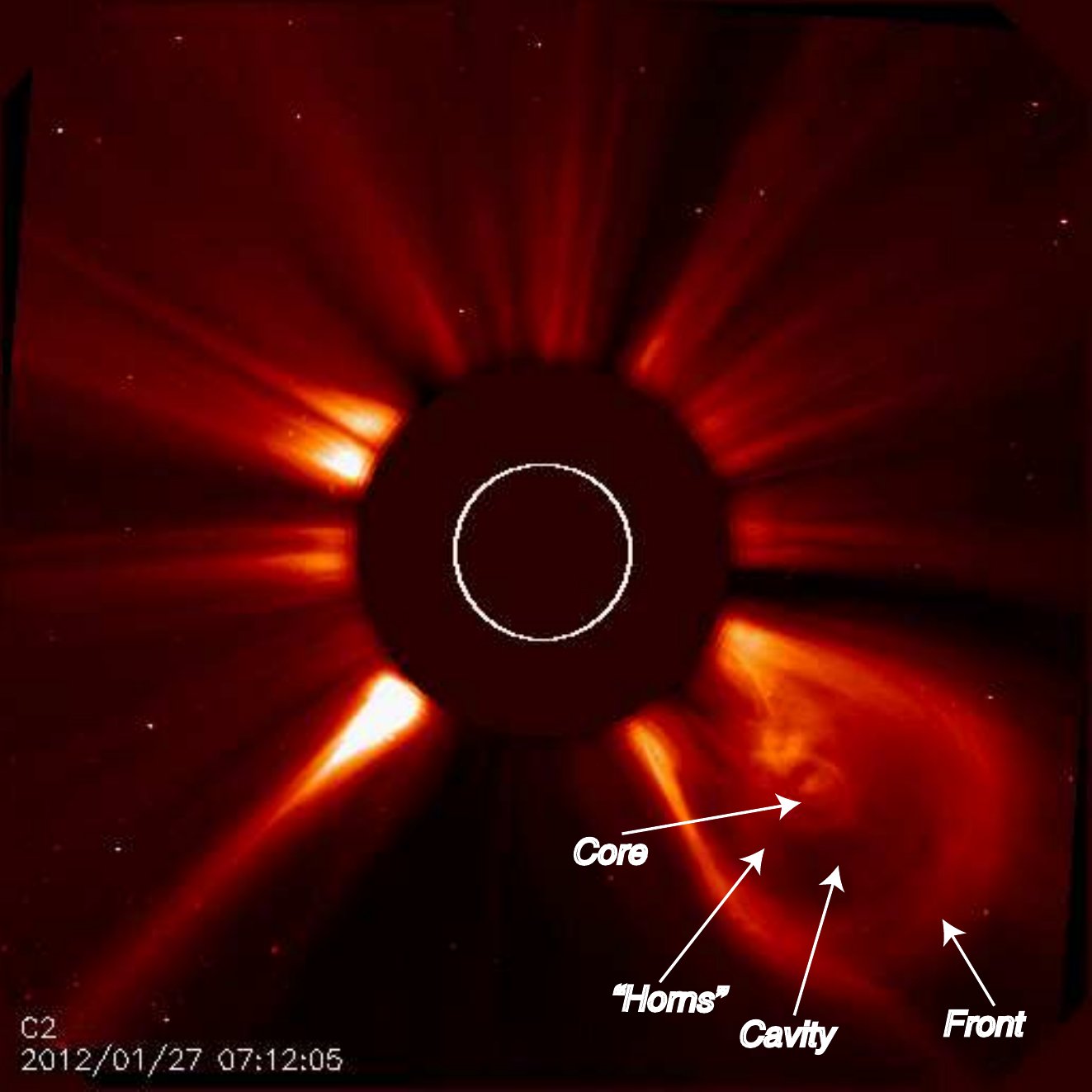}
}
\caption{A typical 3-part CME as it appeared on LASCO/C2 on January
  27, 2012. The three
  components are identified on this snapshot image. The full movie is
  available online. }\label{fig:3part}
\end{figure}

As we noted, the 3-part morphology was identified since the early
coronagraph observations. The prototypical event is an event similar
to the CME in Figure~\ref{fig:3part}. All three components can be
readily identified in this snapshot (the movie is available online)
which was constructed by dividing the original image with a long-term
background to remove the effects of the F-corona but to avoid removal
of the ambient electron corona. An inspection of the accompanying
movie reveals that the brightness of the front originates from the
pile-up of the overlying streamer material. The core has sufficient
structure to identify it unambiguously to the pre-eruption prominence
(we will not discriminate between the terms `filament' and
`prominence' here since they both refer to the same structure). Note
that the cavity, while not completely devoid of plasma, does contain
less electrons (it is less bright) than its surroundings. These aspects have been noted before. The question here is where is
the evidence that the cavity is (or contains) a flux rope like
structure?

Let us focus on the concave upward features labeled as `horns'. They
seem to originate within the core and to outline the extent of the
cavity. Such configuration is consistent with models of prominence
suspension at the bottom of a coronal flux rope cavity
\cite{1995ApJ...443..818L}. \inlinecite{2000SoPh..194..371P} commented
on the appearance of these 'horns' in EIT images before the eruption
as an indication of the formation of the fluxrope which subsequently
erupted and they also noted that the prominence lay at the trailing
edge of the CME. Similar structures were observed by
\inlinecite{1999ApJ...512..484W} and \inlinecite{1999ApJ...516..465D}
and interpreted in a similar way as direct evidence of the FR nature
of the CME cavity.

\begin{figure}
\centerline{
\includegraphics[width=\textwidth]{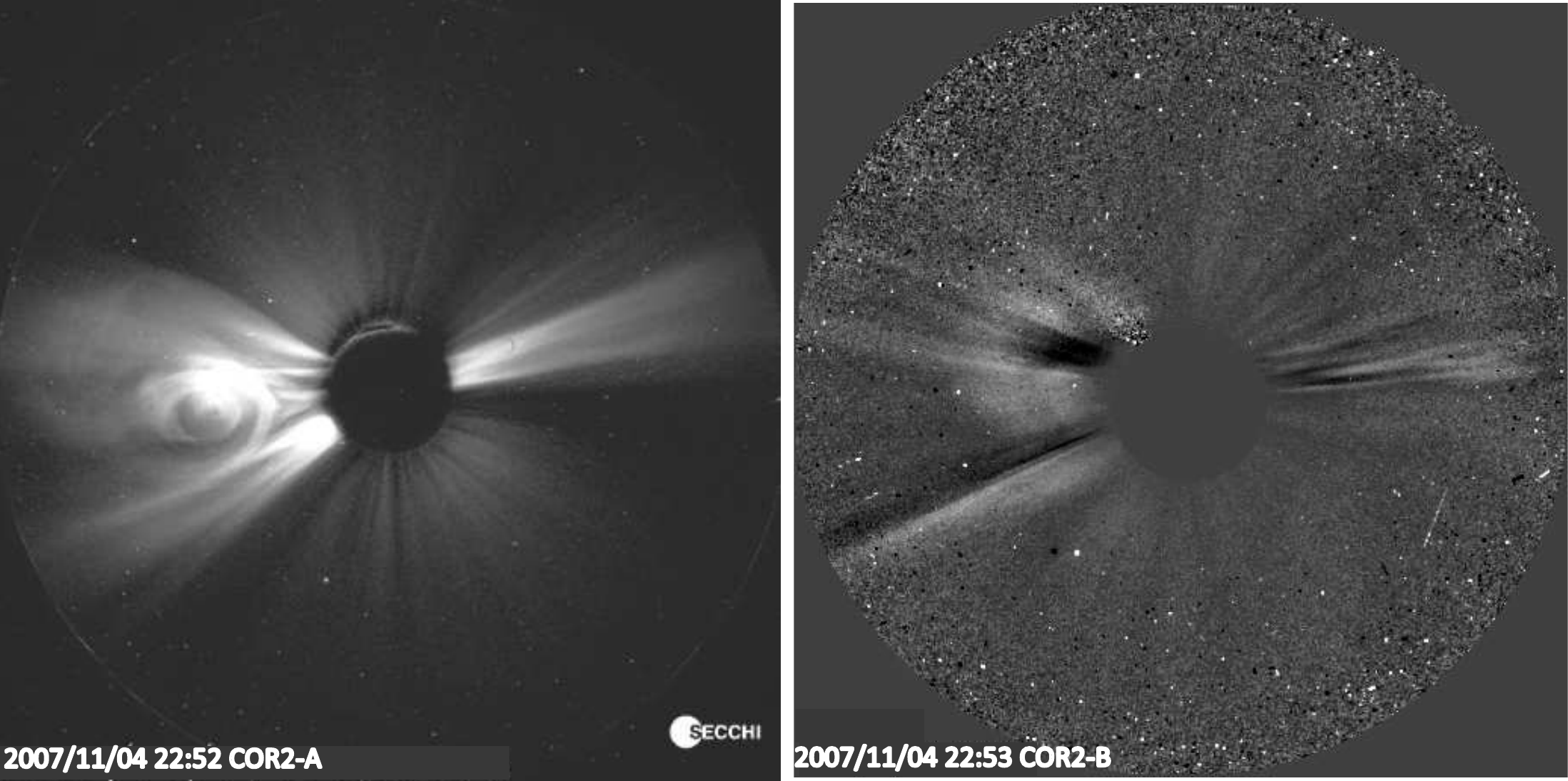}
}
\caption{\textsl{Left\/}: An exceptionally clear example of flux rope
  structure entrained within a CME observed by SECCHI//COR2-A on
  November 11, 2007. There is no 'cavity' in this case since emission
  from the dense flux rope fills that region. Multiple striations
  including a tip at the backend of the structure can be
  discerned. The lack of a bright front is likely due to the low speed
  and high starting height for this event. \textsl{Right\/}:
  Simultaneous image of the same event from COR2-B. The CME appears as
  a partial halo and a base difference image is used to enhance the
  faint emission. There is no evidence of FR structure from this
  viewpoint.  }\label{fig:fluxrope}
\end{figure}

Although these features have been observed in many events since, their
FR association does not seem to be widely recognized. This may be
because the low densities within the cavity do not permit an easy
visualization of the FR structure when only the lower part (the
`horns') are illuminated. The missing `link' would be a 3-part CME
where the cavity would be filled with sufficient amount of plasma to
illuminate the full volume and structure of the
FR. Figure~\ref{fig:fluxrope} shows such 
an example. The event is
associated with a slow eruption of a quiet Sun prominence from the
northern hemisphere. The last traces of 304 \AA\ disappear from the
Extreme Ultraviolet Imager (EUVI)-A field of view at 1:56~UT on November, 4. This is by far the
clearest detection of an FR within a CME despite observations of
thousands of events with LASCO. We believe that the rarity of such
detections is due to four reasons: (i) the clearest signatures will
appear at middle corona heights (say, $>7$ 
R$_{\odot}$) where the
background streamer emission is weaker and the CME has finished
evolving \cite{Vourlidas10}. (ii) the lower spatial resolution of the
LASCO/C3 coronagraph (it is about 
4$\times$ coarser than the LASCO/C2) washes
out some of the fine scale detail. (iii) the FR must have a large
size along the line of sight (LOS) and (iv) the FR must be oriented
almost exactly perpendicular to the POS to produce bright emission
throughout the cavity.  The highly structured core of the CME in the
COR2-A image has an almost identical appearance in the EUVI-A 195~\AA\
images (not shown here) which suggests that most of the core material
was at coronal temperatures ($\sim$1.4~MK) erasing thus any obvious
connections to the prominence (see \inlinecite{2009ApJ...701..283R} for a
very similar example). We will return to this point in
Section~\ref{sec:prom}.

Even such clear observations would not convince probably the skeptics
that the cavity and FR are the same structure. The images are 2D
projections on the POS leaving some room for misinterpretation. A
comparison to theoretical predictions is therefore required. The high
sensitivity of the LASCO images enabled the first opportunity for a
detailed comparison between observations and theroreticaly-derived 2D
FR structures (\opencite{1997ApJ...490L.191C},
\citeyear{2000ApJ...533..481C}). \inlinecite{2006ApJ...652.1740K}
extended these comparisons to a larger sample of LASCO `3-part'-CMEs
in an effort to extract some 3D information (aspect ratio,
eccentricity) of the FR. \inlinecite{2007ApJ...657..559K} attempted to
answer the same question as us by comparing statistical
distributions of the width and the rate of occurrence of
concave-upward structures (essentially `horns' seen in visible light)
from observations to synthetic FR images with satisfactory
agreement. Extensive measurements of the geometric properties of many
3-part CMEs led \inlinecite{2004A&A...422..307C} to conclude that
3-part CMEs were not simply spherical bubbles but structures elongated
along the axis parallel to photospheric neutral line in their source
regions. These results were subsequently confirmed by forward modeling
methods which demonstrated that a 3D geometric representation of a FR-like shape
could account for the observed CME density envelopes and shapes
\cite{2006ApJ...652..763T}. This issue has now been consolidated with
the successful application of FR-like geometric structures on
stereoscopic observations from STEREO
(\opencite{2009SoPh..256..111T}, \citeyear{2011JASTP..73.1156T}; 
\opencite{2011JASTP..73.1201R}; \opencite{2011ApJ...729...70W}).
There should be little doubt, therefore, that `3-part'-CMEs are indeed
systems of ejected FRs where the cavity is the actual FR.

Our examples, and the events exhibiting clear 3-part structures in
general, must lie close to the POS to provide an edge-on view of
the FR. But CMEs occur in all longitudes.  Occasionally, a `3-part'-CME
will be observed along the Sun-Earth line. How can we then tell
whether a halo CME has a 3-part structure and how can we identify
the erupted FR against the backdrop of deflected streamers and
material outflows?  For this, we first need to identify the signatures
of the other erupting structures starting with the shock.

\section{Where is the Shock? The 5-part CME} \label{sec:shock} 

It is common knowledge that sudden changes in plasma motion will
launch compressive waves through the medium. If the motion occurs
faster than the characteristic speed of the medium (ie., sound speed
for an unmagnetized plasma or fast-mode speed for a magnetized plasma)
the waves will then steepen into shock waves. In either case, the
propagaing wave will manifest itself as a propagating compression of
density (and magnetic field where applicable). In our case, the
propagating FR will generate a plasma wave which may look as another
propagating front in a coronagraph (or EUV) image sequence. Depending
on various factors, such as the impulsiveness and starting height of
the event, and the magnetic and plasma configuration in the ambient
corona, the density compression of the wave front could become strong
enough to be detected \cite{2003ApJ...598.1392V}. Because waves (and
shocks) are an intrinsic component of any eruption, their coronagraphic
signatures have been the subject of debate since the first CME
observations (see \inlinecite{2009AIPC.1183..139V} for a historical
discussion). It was generally accepted that distant streamer
deflections were a reliable, but indirect, proxy for these waves
\cite{1974JGR....79.4581G,2000JGR...105.5081S}. 

The first identification of the density enhancement from a CME-driven
shock was reported by \inlinecite{2003ApJ...598.1392V} thanks to the
high sensitivity of the LASCO observations. Such signatures are now
commonly reported in the literature
\cite{2006SoPh..239..277Y,2009ApJ...693..267O,2009AIPC.1183..139V,2009SoPh..259..227G,2010ApJ...720..130B,2012AIPC.1183..139V,2012ApJ...746..118K}. We
now know that the CME-induced waves can be detected in coronagraph
images, that they are faint and, that they are located ahead of the FR
front. So it should be straightforward to identify them in any image
(assuming there is a reasonable expectation of a wave occurrence due
to the speed of the CME, for example). Because these wave signatures
are faint, the best approach is to use calibrated, excess mass
images (to remove effects such as vignetting, background streamers,
etc) and display them with high contrast.

\begin{figure}
\centerline{
\includegraphics[width=0.9\textwidth]{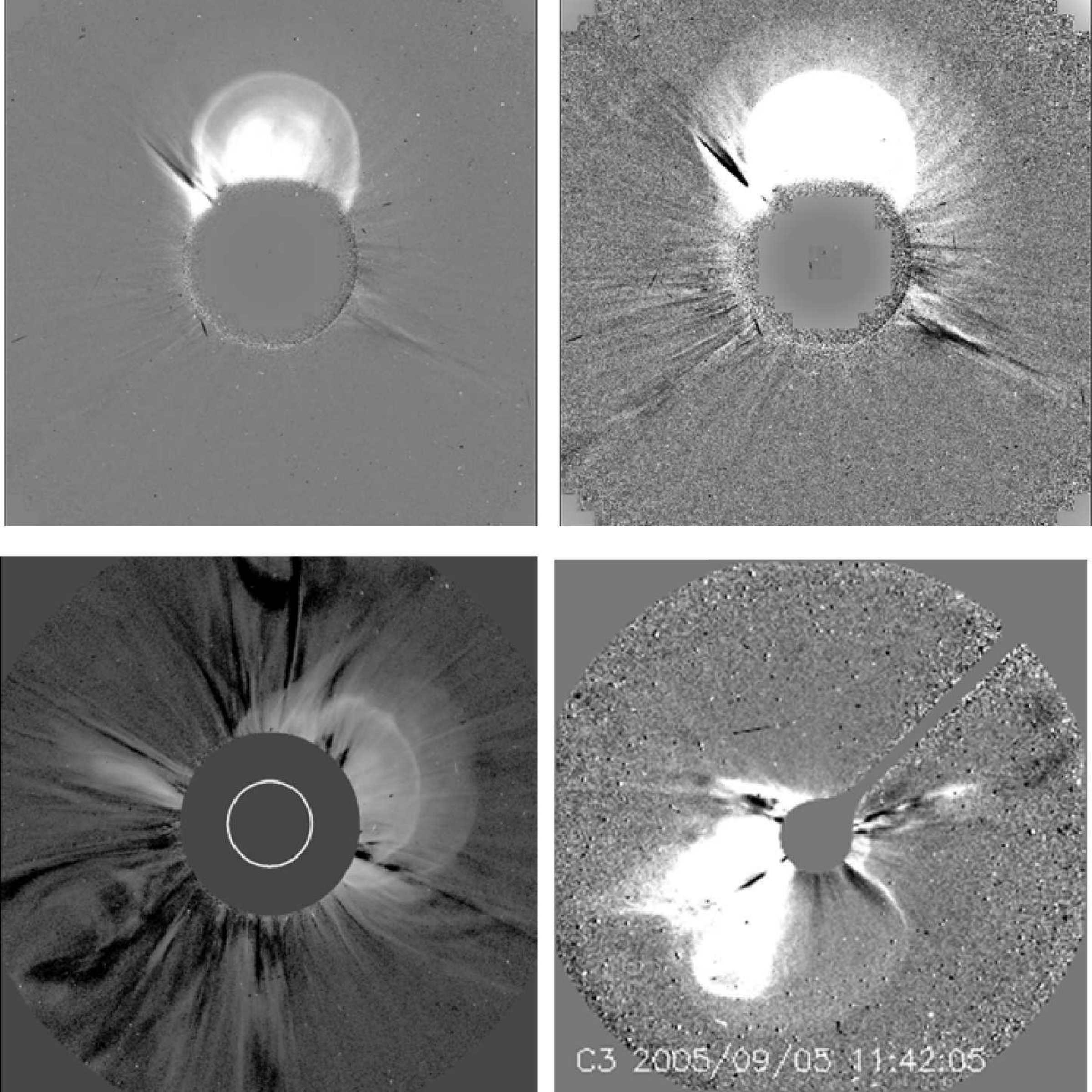}
}
\caption{\textsl{Upper left\/}: An excess mass image of a CME in the
  LASCO/C2 field of view. The event can be easily classified as a
  3-part CME. A deflected streamer is located to the east without an
  apparent connection to the CME. \textsl{Upper right\/}: The same
  image displayed with a higher contrast ratio. A faint arch front
  terminating at the eastern deflected streamer is now
  visible. Indications of several other deflected streamers (or
  sub-streamer structures) can be seen. \textsl{Bottom panels\/}:
  Other examples of shock fronts ahead of CMEs in C2 (left) and C3
  (right). }\label{fig:shock_comp} 
\end{figure}

An illustrative example is shown in the upper panels of
Figure~\ref{fig:shock_comp} where the same frame from a fast CME is
shown with two different contrast ratios. On the left, the CME has the
classical 3-part appearance with a very clear loop front. On the
right, the higher contrast ratio allows to see a fainter front ahead
which extends around the bright loop front and connects to the
deflected streamer on the eastern flank. A series of other deflected
streamers (or more likely substreamer structures) can be seen as
radial striations occupying position angles from the deflected
streamer, around the CME to the western equator. The faint front
appears to be the outer envelope of these deflected streamers
consistent with being a wave driven by the CME and propagating within
a large-scale streamer. Two more examples of such fronts are shown in
the bottom panels of Figure~\ref{fig:shock_comp}. On the left, a C2
image from a CME on June 11, 2000 shows a bright filamentary front (a
Loop CME) preceeded by an extensive faint front connecting to a
deflected streamer in the north. On the right, a C3 image shows again
a faint front terminating at a deflected streamer but this time the
front extends only to one side of the CME. No fronts, and tellingly no
loop front either, are seen along the northern CME flank.

These images serve to illustrate our earlier point that the
detectability of these 
shock fronts is highly dependent on the
sensitivity of the observations. It is thus unsurprising (in
retrospective) that such features have eluded detection in the pre-LASCO coronagraph experiments which lacked CCD
detectors, large fields of view, and long-term uninterrupted
observations. For many events, only the bright loop would be detected
(Figure~\ref{fig:shock_comp}, top left) thus only allowing rather
indirect and ambiguous arguments on the existence of a shock
\cite{1987JGR....92.1049S}.

A plausible criticism that may arise from our interpretation of these
images is how can we be sure that the faint front is indeed related to
density pile-up at a wave front and is not simply ejected material,
i.e., coronal loops moving ahead of the CME in direct analogy to the
bright front ahead of the cavity. This can be best answered by careful
inspection of excess mass movies of these events. If the front in
question is due to ejected material then a depletion should form
behind it as it does behind the CME proper. If the front is caused by
a wave, then the enhancement is due to density compression and not
material transport, therefore no depletion should occur. The latter
is the observed behavior for the events in
Figure~\ref{fig:shock_comp}. MHD simulations have provided further
support for this interpretation by matching the location and density
compression ratio between observations and model
\cite{2003ApJ...598.1392V,2008ApJ...684.1448M}. Recently, similar
fronts have been detected in high cadence EUV images
\cite{2011ApJ...738..160M,2012ApJ...745L...5C}.

Returning to the white light morphology, the identification of the
faint front ahead of the bright loop-like CME front simplifies greatly
the interpretation of CME images. The bright loop is the pile-up
of material at the outer boundary of the erupted FR (the cavity) and
hence it is bright while the outer front originates from a temporary
compression of the ambient plasma as the wave (or shock) propagates
through and hence is much fainter (see \inlinecite{2009ApJ...693..267O} for density
profiles of these structures). Our examination of thousands of CME
images (Section~\ref{sec:stats}) reveals that the ``faint front followed by
a bright loop'' is a common occurence and it can constitute a reliable
signature for the identfication of \textsl{both the shock and FR
  fronts\/} in the images. An important benefit from this
identification is a better interpretation of the structures in images of
halo CMEs. 

\subsection{Halo CMEs}
There is no physical reason to expect that halo CMEs are a differenct
class of CMEs. `Loop' and `3-part'-CMEs should occur as halos and their
FR should appear on the images...somewhere. The common approach has
been to identify the outer envelope of the halo with the FR and
proceed to fit it with a circular or elliptical cone models to extract
kinematic parameters
\cite{1982ApJ...263L.101H,2002JGRA..107.1223Z,2004JGRA..10903109X,2005JGRA..11008103X}. However, this approach is inconsistent with our theoretical understanding of
FRs as more or less cylindrical structures, elongated along their
axis. It is also inconsistent with the analyses of the CME projection
effects \cite{2004A&A...422..307C,2009SoPh..256..111T}. Our discussion
above solves this problem. To identify the FR in a halo CME image, we
have to look for evidence of the `two-front' morphology or of the
bright loop structure alone when the CME is not fast enough to expect
significant pileup at its wave front. Indeed, these structures are
visible in the majority (if not all) of halo CMEs.

\begin{figure}
\centerline{
\includegraphics[width=\textwidth]{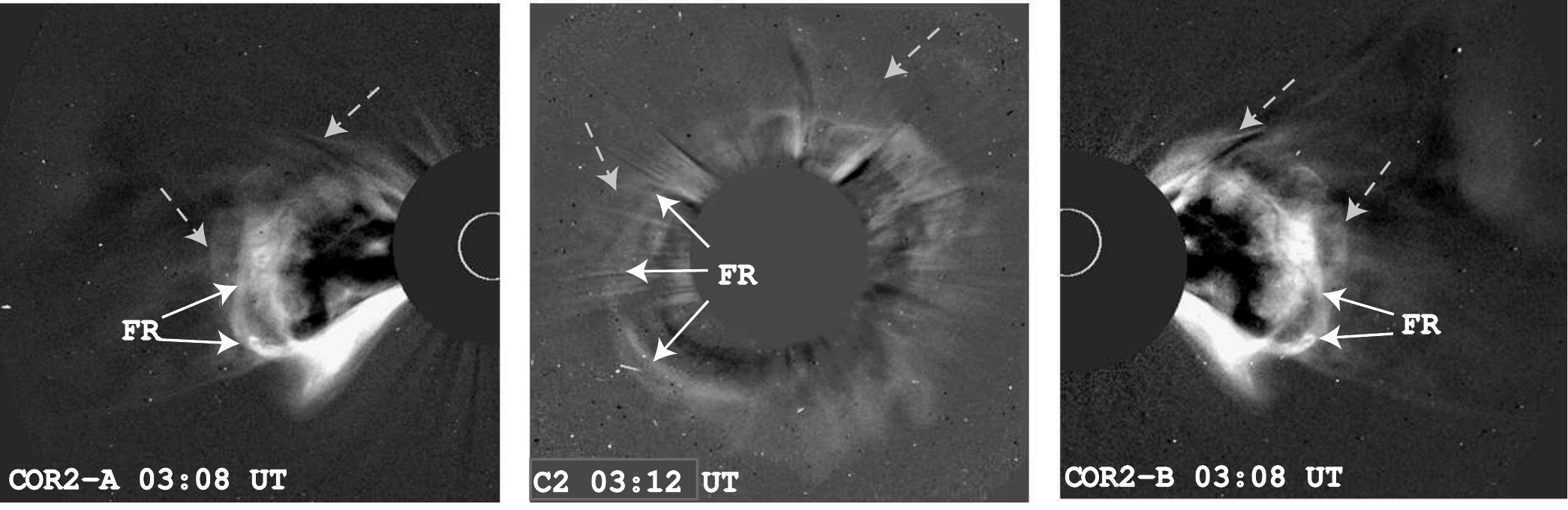}
}
\caption{A demonstration of the appearance of `two-front' morphology
  in a weak halo CME as viewed from LASCO/C2 (center). The CME appears
  as a `loop'-CME in the SECCHI COR2-A (left) and B (right)
  coronagraphs. All three images are taken nearly simultaneously. The
  pile-up at the FR edge is marked by solid white arrows. The edge of
  the much fainter wave is marked by the dashed grey arrows. The
  existence of a shock is likely since the CME is propagating at $\sim
  750$ km s$^{-1}$ in the LASCO field. The event occurred on February,
  15, 2011 in association with an X-class flare. For similar examples
  see Vourlidas and Ontiveros (2009), Vourlidas and Bemporad (2012).}
\label{fig:halo_ex}
\end{figure}

We picked a recent halo CME as an example
(Figure~\ref{fig:halo_ex}). The event, which occurred on February 15,
2011, was the most symmetrical halo of the current solar cycle. It was
associated with a large X-class solar flare, metric and decimetric
Type-II emissions, and an EUV wave. We chose an event with relatively
weak halo emission to demonstrate the robusteness of the feature
detection. Much clearer examples are presented in
\inlinecite{2012AIPC.1183..139V}. See also
\inlinecite{2009AIPC.1183..139V} and Figure~\ref{fig:comp} (bottom
left) for a single LASCO view.
Returning to Figure~\ref{fig:halo_ex}, we see that the halo CME
appears as a regular `loop', or even `3-part'-CME in the COR2-A and B
fields of view. Note that all three images were taken nearly
simultaneously and are differenced from a pre-event image. The FR
boundary is readily identified in the COR2 images as a bright loop
structure (marked by the white arrows and the label `FR'). The same
structure appears as a (fainter) loop in the C2 image. A still fainter
front can be seen ahead of the loop. A wave compression, and possibly
a shock, is expected in this event given its LASCO speed of $\sim750$
km s$^{-1}$ and the Type-II radio emission. The wave is more difficult
to discern in the side views from COR2 (dashed gray arrows) because
the CME is projected against, and propagating though, a background
corona disturbed by an earlier event. Even the streamer deflection and
the wave associated with it can be detected, albeit barely, in the
north (topmost gray arrow). The rugged set of features along the
southern-southeastern part of the halo originate in the bright
southern streamer, seen in the COR2 images. 

This exercise shows that, with a little effort, we can identify the
origins of the various features in a halo CME image and delineate the
boundaries of the ejected FR with some precision. The important point
here is that we do not have to rely on simplistic, rough
approximations for the envelop of the CME. These would lead to
imprecise measurements of the CME speed, size, and orientation with
correponding implications for Space Weather predictions.

\section{Theoretical Support: Synthetic Images from MHD Simulations} \label{sec:models}

Thus far, we have only used coronagraph images to support our
interpretation of CMEs as 3-part (5-part when a wave front appears)
structures resulting from the expulsion of a magnetic FR from the
Sun. 
We now turn to a numerical MHD simulation to determine which aspects
of the coronal signatures identified in the LASCO images can be
produced by an erupting three-dimensional FR.
We analyze results obtained with an Adaptively Refined MHD Solver
(ARMS; \opencite{DeVore2008}) simulation of the `magnetic breakout'
CME initiation mechanism (\opencite{1999ApJ...510..485A};
\opencite{2004ApJ...617..589L}, \citeyear{Lynch2008}) and the
subsequent FR-CME propagation in the low corona to construct synthetic
coronagraph images we can directly compare to observations.

\subsection{Description of the MHD Simulations and Eruption Overview}

\begin{figure}
\centerline{
\includegraphics[width=1.0\textwidth]{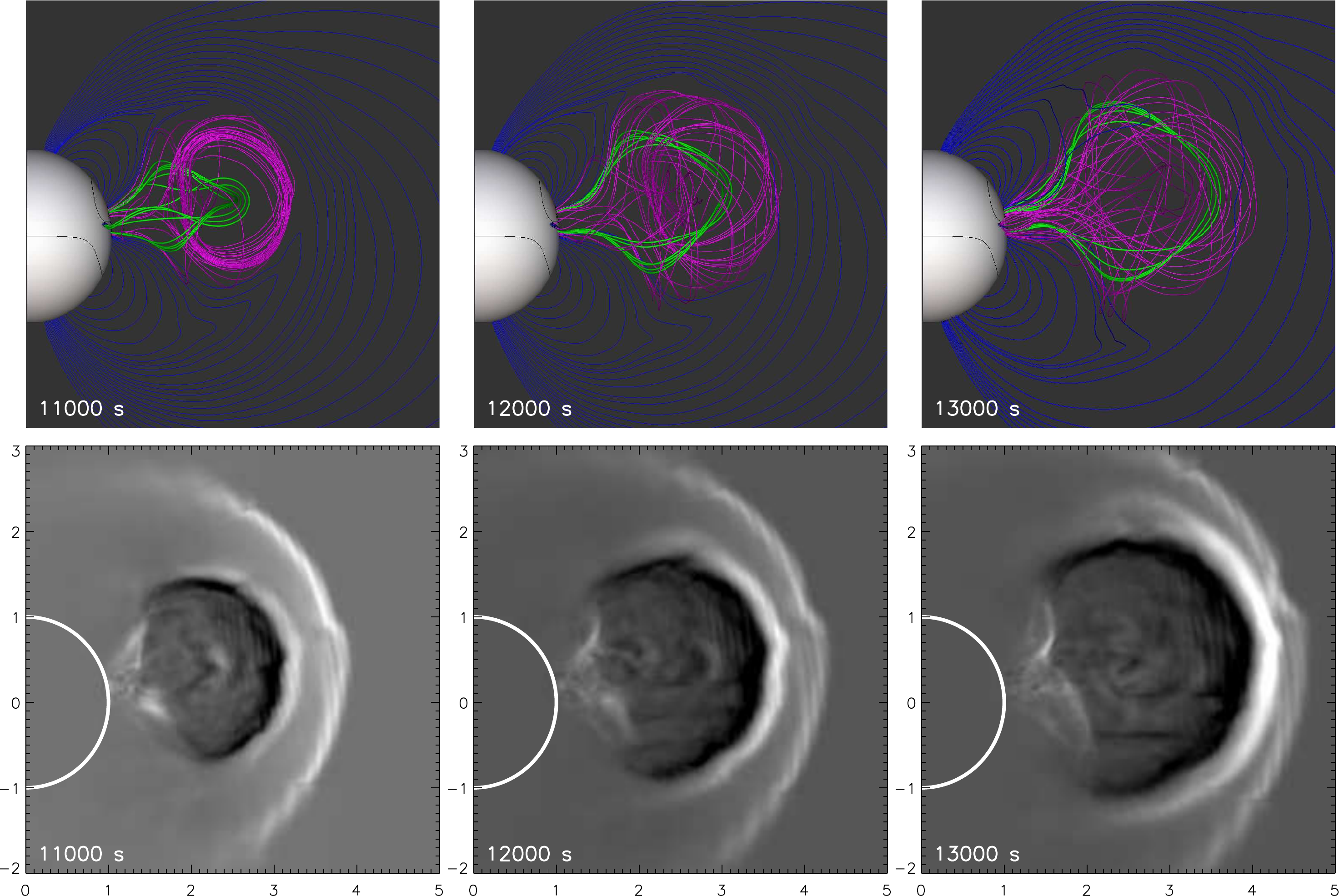}
}
\caption{\textsl{Top row}: MHD simulation results of a breakout CME
eruption in the coronagraph field of view for a limb event. The FR
ejecta fieldlines are green, magenta for the sheared field core and
reconnection generated twist flux. \textsl{Bottom row}: Synthetic
running-difference (RD) coronagraph images constructed from the
simulation's density evolution.}\label{fig:evolution}
\end{figure}

The ARMS simulation data analyzed herein comes from the `Left-Handed'
CME eruption described by \inlinecite{Lynch2009} in a
fully 3-dimensional, globally multipolar magnetic field configuration.
The solar atmosphere is initally in gravitationally stratified
equilibrium with spherically symmetric density, pressure, and
temperature profiles given in \inlinecite{Lynch2008}. The maximum
field strengths in the AR are $\pm40$~G which, while lower than
observed values by anywhere from 10-100, yield a low-$\beta$ plasma
in the CME source region ($\beta \sim$10$^{-3}$) and throughout the
computational domain. Thus, the ARMS simulation data provide a
physically valid, albeit idealized, representation of the
magnetically-driven eruption process.
Here, we briefly review the phases of the moderate speed
\inlinecite{Lynch2009} breakout CME eruption:

\begin{enumerate} 
\item \textsl{Energization} ($ 0 \le t \le 10000$~s): Surface
shearing flows are applied adjacent to the polarity inversion line
(PIL) of the active region (AR) resulting in the gradual accumulation
of magnetic energy ($E_{\rm M} \sim$10$^{31}$~ergs) as the low-lying,
strong AR fields are stressed. This sheared field component parallel
to the AR PIL will become the FR-CME axial field.
\item \textsl{Breakout Reconnection} ($t \gtrsim 5000$~s): As the
sheared portion of the AR flux expands, the overlying coronal null
point becomes distorted and flattened, forming a current sheet at
the separatrix between the AR and background flux system. Continued
expansion compresses the current sheet and drives magnetic reconnection
which transfers overlying restraining background flux out of the
way of the expanding stressed field which, in turn, increases
expansion, and drives more reconnection in a runaway positive
feedback scenario. The breakout reconnection facilitated expansion
shows up as a smooth increase in kinetic energy to $E_{\rm
K}\sim$3$\times$10$^{29}$~ergs.

\item \textsl{Eruptive Flare Reconnection} ($t \gtrsim 10000$~s):
The runaway sheared arcade expansion drive the formation and
elongation of a radial current sheet above the PIL leading to the
start of flare reconnection in the shear channel. The eruptive flare
reconnection rapidly releases stored magnetic energy ($\Delta E_{\rm
M} \sim 7\times10^{30}$~ergs) through the magnetic reconfiguration
and formation of flare loop arcades, supplies material and momentum
to the ejecta via strong reconnection jet outflow ($E_{\rm K}$
increases to $1.05\times10^{30}$~ergs), and, in the breakout model,
\textsl{creates} the magnetic flux rope during the eruption process
by generating highly twisted flux surrounding the erupting
sheared field core.
\end{enumerate} 
The top row of Figure~\ref{fig:evolution} plots representative
magnetic fieldlines for $t=\{$11000, 12000, 13000$\}$~seconds during the
CME eruption. Fieldlines representing the FR sheared field core are
plotted in green, the reconnection-created FR CME twist component
in magenta, and the background field in dark blue.


%
\textbf{There are two primary challenges associated with MHD modeling of
very fast CMEs and their subsequent shock generation in the low
corona. The first is a correct description of the thermodynamics,
field and plasma structure of the steady-state background solar
wind. The lack of a single, widely-accepted theory for coronal
heating means, in practice, every simulation relies on idealized,
parametrized heating terms and calculates the resulting solar wind
and open field structure from the balance of forces. The second
major modeling challenge is overcoming the computational limitations
associated with the magnetic field strength in CME source regions.
The MHD numerical timestep is limited by the Alfven speed which
makes the temporal evolution of kilogauss fields that are routinely
observed in large active regions prohibitively expensive.  Despite
these model limitations, numerical MHD simulations are becoming
increasingly sophisticated and capable.  For example, the field
strengths and self-consistent pre-eruption energization used by
\inlinecite{Lynch2008} and \inlinecite{2008AIPC.1039..286R} were
sufficient to initiate CMEs with eruption speeds on the order of
1200--1400~km/s. In simulations that bypass the difficulty of the
pre-eruption evolution, fast eruptions can be generated with CME
speeds $>$2000~km/s, drive shock formation as low as
$\sim$1.6~$R_\odot$, and can produce complex white-light structures
in synthetic coronagraph and HI images
\cite{2008AIPC.1039..286R,2008ApJ...684.1448M,2011ApJ...738..127L}.}

%
%

\subsection{Comparison Between Synthetic and LASCO CME Images}
%
\begin{figure}
\centerline{
\includegraphics[width=1.0\textwidth]{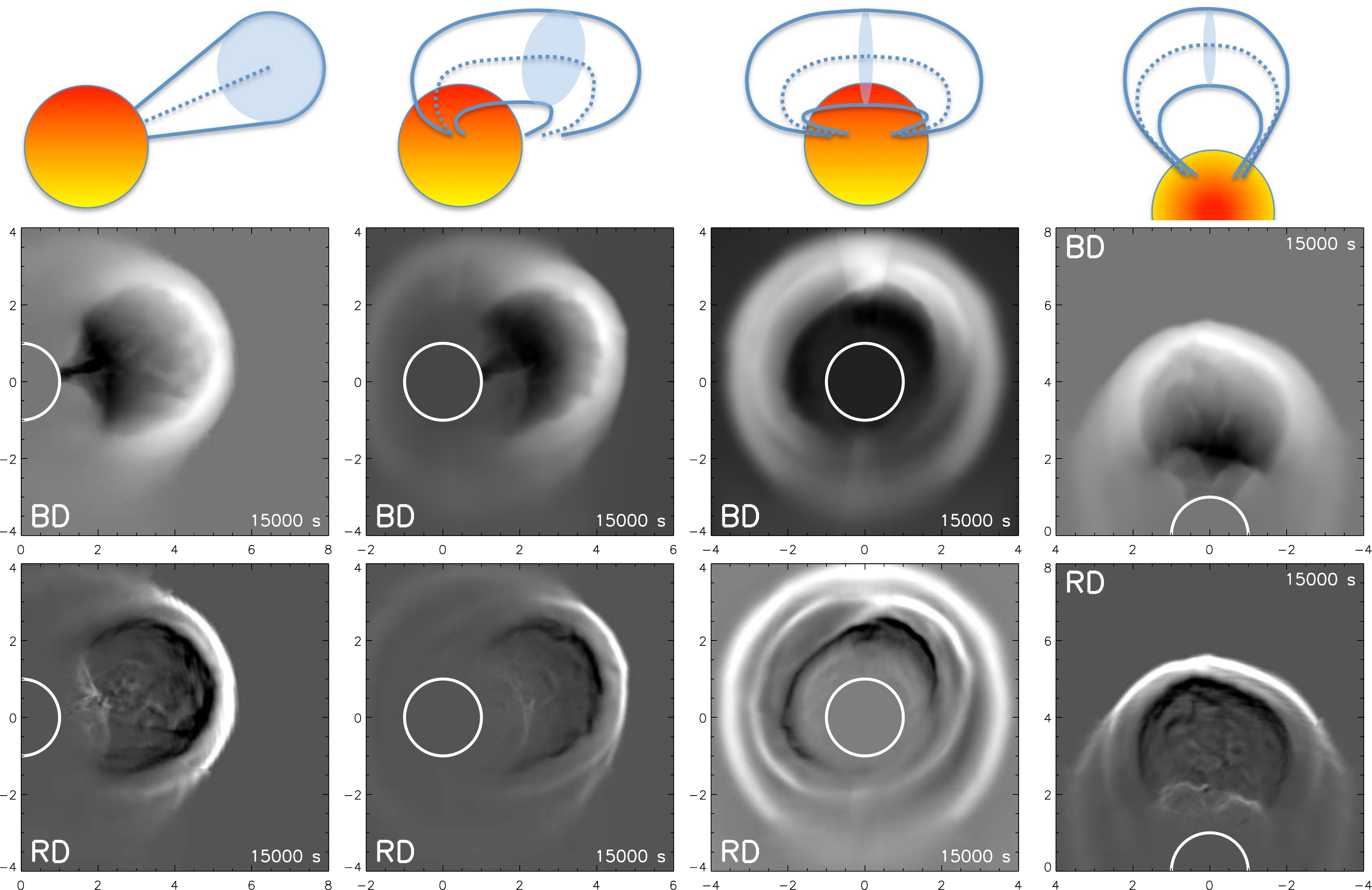}
}
\caption{\textsl{Top row}: Schematic images depicting the 3D FR
  orientation with respect to the synthetic image POS. The
  \textsl{middle}, \textsl{bottom} rows plot the synthetic
  base-difference (BD), running-difference (RD) images,
  respectively. Both the bright CME front and the CME-driven
  shock/expansion wave can be seen in these images.}\label{fig:bd_rd}
\end{figure}
%

To construct synthetic coronagraph images, we use a version of the
SolarSoft routine {\tt eltheory.pro} to calculate the total brightness
in a regular 2D Cartesean array of lines of sight that 
sample the spherical 3D MHD density data at every simulation output time. 
From these synthetic total brightness images we then construct
base-difference (BD) images as $B(t)-B(0)$ and running-difference (RD)
images as $B(t)-B(t-\Delta t)$. The temporal
cadence, $\Delta t$, of our simulated data is 250~s.
The bottom row of Figure~\ref{fig:evolution} plots the synthetic
RD images such that the radial propagation of the center of the FR
CME lies exactly in the RD image plane of the sky (POS) corresponding
to the viewpoint of the 3D fieldline visualization.

In Figure~\ref{fig:bd_rd} we have constructed a series of viewpoint
orientations to examine morphological features of the synthetic
coronagraph images.  The top row indicates schematically the CME
orientation with respect to the image POS and the middle, bottom
rows plot the corresponding BD, RD images respectively. From left
to right, the angles between the radial propagation of the center of
the FR CME and the image POS are $\{$0, 45, 90, 0$\}$~degrees, with
the fourth column representing a vantage point from the North solar
pole looking down on the eruption.

The shock front, the bright CME front, and the FR cavity are each
clearly seen in either the RD or BD images. The CME front morphology
varies between a loop-like CME (within $45^\circ$ from the limb)
to a halo CME (at $90^\circ$ from the limb) similarly to the actual
observations.
 
While the idealized MHD simulation produces both the shock
front/expansion wave and the bright leading edge of the ejecta ahead
of the FR CME driver, there is no corresponding high density plug
of material associated with the FR core. This limitation was also
present in the axisymmetric models (e.g., \opencite{2004ApJ...617..589L})
and is due largely to our simplified pre-eruption coronal density
distribution which does not include prominence material along the
low-lying sheared field or the enhanced densities associated with
either ARs or that would arise in a closed-field streamer belt
geometry. Our simplified model background results in two main
consequences.
First, without dense material tracing the topology of the FR CME core, it is
difficult to distinguish between the edge-on view (Figure~\ref{fig:bd_rd}
first column, looking at the FR cross-section; compare to Figure~\ref{fig:3part}) and the top-down view
(Figure~\ref{fig:bd_rd} fourth column, FR axis lies in the image
POS; compare to Figure~\ref{fig:shock_comp}, bottom left) in the synthetic images. 
Second, the relative brightness of the CME leading edge and the
shock/expansion wave do not have the same ratio as commonly observed
in the coronagraph images (which, as discussed earlier, require significant
contrast enhancement).
Furthermore, the lack of a background with coronal streamer structures
in various locations does not allow us to compare streamer deflections
or the effects of coronal hole locations in this particular simulation,
although these issues are an area of active research (see, e.g.,
\opencite{2011ApJ...738..127L}; \opencite{2012ApJ...744...66Z}).
 
We also note that the synthetic shock appears very close to the CME
leading edge (the driver) which is not the case for the LASCO images
we have presented so far. This is an evolutionary effect, however,
as Figure~\ref{fig:evolution} has already shown.
In our simulations, the shock is initially clearly ahead of the CME
which catches up to it within $\sim$2000 sec. Of course, different
speed profiles and ambient coronal configurations will result in
different standoff distances. Again, our comparison here is not an
attempt to model a specific CME event with a realistic background
density distribution, but rather to present an idealized general case
of the appearance of a generic shock-driving FR-CME in a coronagraph
field of view.

\begin{figure}
\centerline{
\includegraphics[width=0.9\textwidth]{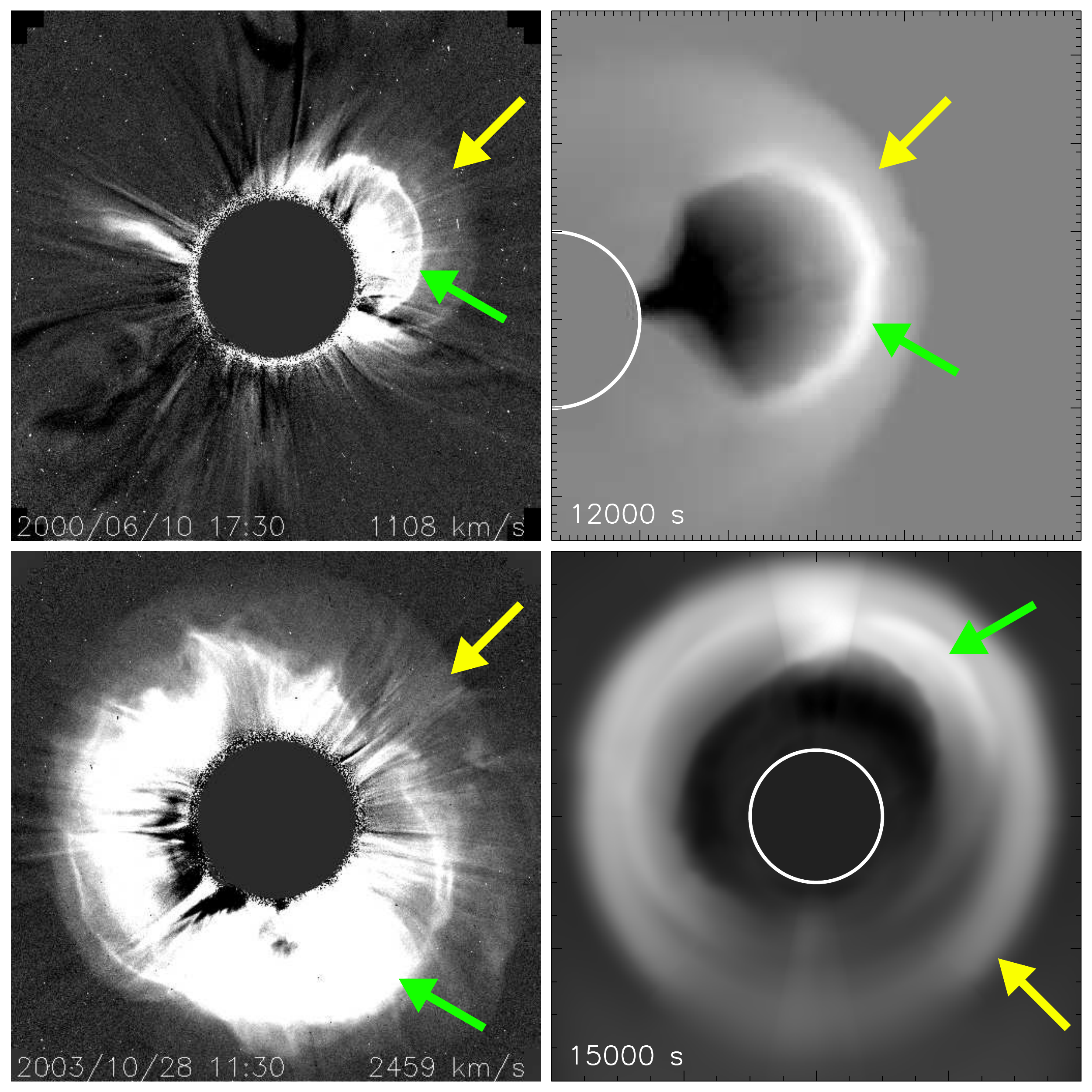}
}
\caption{Comparison between LASCO excess mass images (left column)
and synthetic base-difference images from our MHD simulations (right column).
There is a very good correspondance between observed and simulated
structures, despite the idealized nature of the simulation.  The
yellow arrows denote the shock/expansion wave front and the green
arrows denote the CME leading edge density enhancement. The projected speeds for the LASCO CMEs are also shown.}\label{fig:comp}
\end{figure}

It is precisely the generality of our simulation that makes the
comparison to two LASCO events especially striking, as illustrated
in Figure~\ref{fig:comp}. The simulated synthetic BD images (right
column) show the diffuse intensity region leading the bright CME
front which is associated with the boundary of the magnetic FR
structure exactly as we proposed in Sections~\ref{sec:3part} and \ref{sec:shock}.
Yellow arrows denote the shock front, green arrows denote the CME
leading edge. 
Despite not capturing the observed sharpness or intensity of the
CME front, the simulations do show that the faint halo outline in
the LASCO images corresponds to the shock envelope. Therefore, our
5-part CME structure is a completely straight-forward interpretation
and a natural consequence of the eruption of a 3D FR CME.
%
%
%

\section{The Relationship between Ejected Prominence and CME Flux Rope} \label{sec:prom} 

Having discussed the nature of the front and the
cavity, we turn our attention to the core, the last component of 
`3-part'-CMEs. The core has been associated with the erupting prominence
ever since combined H$\alpha$ and coronagraphic observations
demonstrated the colocation of the two structures (e.g.,
\opencite{1975SoPh...45..363H,1985JGR....90..275I}). The fuzziness of the core compared to
the filamentary structure of prominences in H$\alpha$ and
He~I
observations is attributed to the progressive ionization and heating
of the chromosheric plasma within the prominence as the structure
erupted outwards. 

Of the many unclear aspects on the role and evolution of prominence in
eruptive events, there are two that pertain to our discussion
here. Namely, what is the spatial relationship between the core (and
prominence) to the cavity and, why in-situ detections of chromospheric
material within CMEs are so rare.

The first question stems from the early interpretations of the
prominence as a flux rope and its identification with the CME
cavity \cite{1995ApJ...443..818L}. Although prominences exhibit helical
structures when they erupt, the large number of combined observations
with LASCO (in white light) and  EIT (in He~I 304\AA) have shown that
the cavity is not the prominence itself. They have also shown
that the core does not lie in the center of the cavity as was thought
in the past (see Figure~10 in \opencite{1986ApJ...305..920C}). Rather
the prominence lies at the bottom of the cavity, the cool plasma
suspended in the dips of the FR field lines by the balance between
gravity and magnetic tension forces. 

\begin{figure}
\centerline{
\includegraphics[width=0.9\textwidth]{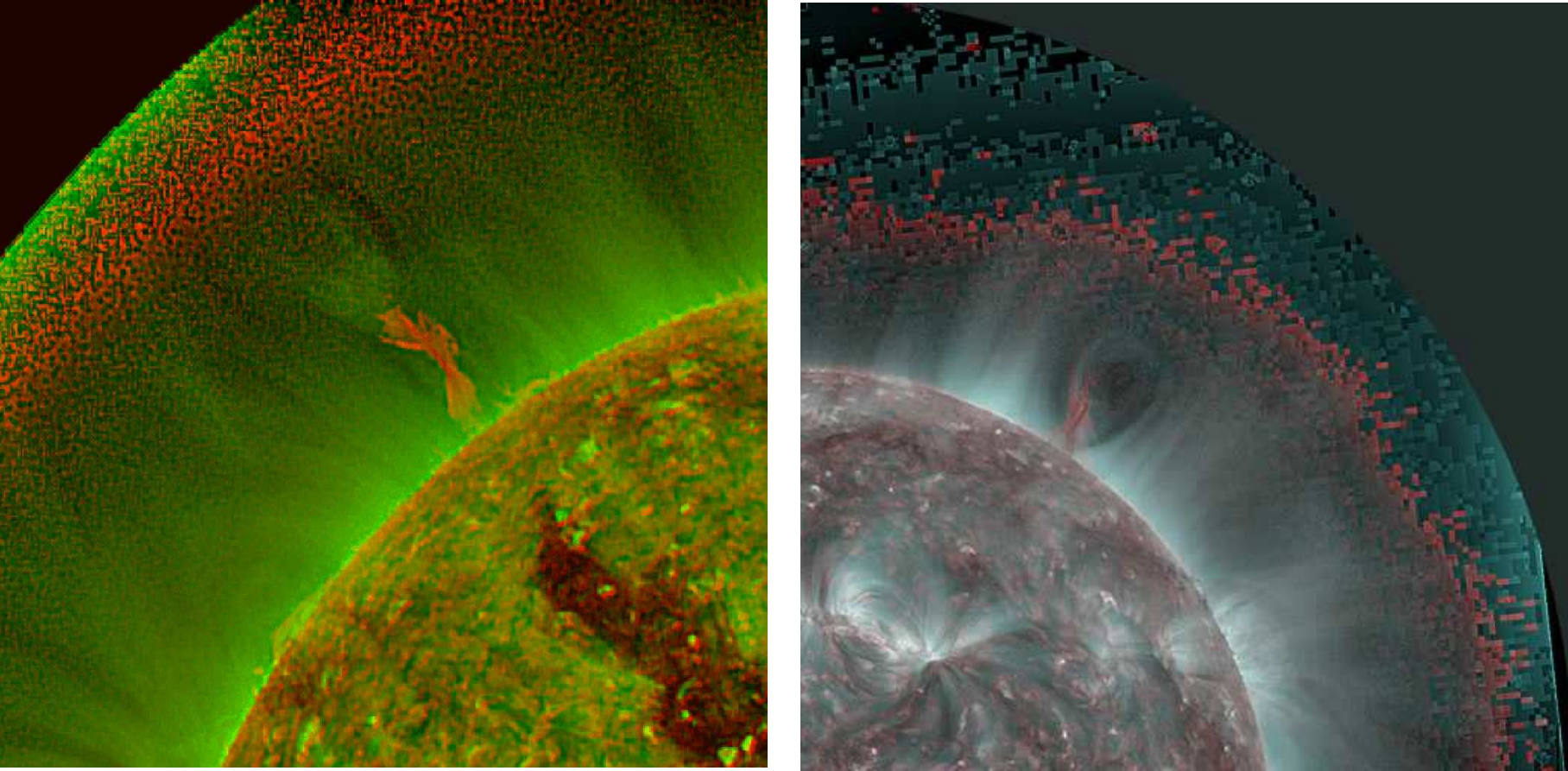}
}
\caption{\textsl{Left\/}: Image of a prominence eruption at the
  northeast limb on June 12, 2010 using a composite of AIA 193\AA (green) and 304\AA
  (red) channels. {Right\/}: Snapshot of a prominence eruption at the
  northwest limb on February 28, 2010 using a composite of EUVI-A
  195\AA (silver) and 304\AA (red) channels. Both images are snapshots from the online
  movies.  
}\label{fig:2color}
\end{figure}

While these concepts are widely accepted thanks to the extensive
observations of quiescent cavities and prominences
\cite{2006ApJ...641..590G}, there seems to be a lingering confusion on
the location and importance of the prominence relative to the erupting
FR or cavity. The high resolution observations from SECCHI and
Atmospheric Imaging Assembly (AIA; \opencite{2012SoPh..275...17L}) can
now put this issue to rest. For example,
\inlinecite{2011A&A...533L...1R} have presented multi-wavelength AIA
observations of a polar crown filament during its early eruption
stages where they image both the cool prominece and the hotter FR. The
high spatial resolution of AIA reveals cool prominence plasma in 304
\AA\ embedded in field lines at the bottom of the FR visible in 193
\AA\ ($\sim1.4$ MK). The 193\AA\ emission has the typical `horn'
morphology we discussed in Section~\ref{sec:3part} and frequently seen
in the EIT images during Cycle 23. To demonstrate how common are these
structures, we present two more examples in Figure~\ref{fig:2color}.

On the left, we show a snapshot from a prominence eruption at the
northeast limb on June 12, 2010 captured by the AIA instrument. To
demonstrate the relative locations of the FR and prominence we
combined the 193\AA\ (gold) and 304\AA\ (red) AIA channels after
enhancing the individual images through our wavelet processing
algorithm \cite{2008ApJ...674.1201S}. The accompanying movie
demonstrates two important aspects of the eruption: (1) the
pre-existing cavity is not actually empty but it is filled with plasma
at coronal temperatures, and (2) most of the cool prominence material
returns to the surface and is not ejected with the rest of the CME.

On the right, we show a
similar observation from SECCHI/EUVI-A taken on February 28,
2010. Here the 195\AA\ images are shown in silver color. We observe
strong circular motions in the center of the cavity during the early
part of the event previously noted by \inlinecite{2010ApJ...719L.181W}
who did not comment on the temperature characteristics on these
motions. Here, we can see that the motions are associated with
extensions of 304\AA\ emission into the center of the cavity which
exhibits a very clear FR morphology. The cool plasma seems to
disappear after its injection until a new extension brings in another
quantity of cool material into the cavity. Hence, the rotations within
the cavity seem to be driven by the episodes of heating of
chromospheric plasma. In contrast to the June 12, 2010 event, the
prominence erupts carrying a significant amount of cool plasma
outwards. No return flows are evident in this event.

These EUV observations are quite clear. The prominence is not the
cavity. It is the core. The core is not located at the center of the
cavity. It is located at the trailing edge. It may, however, appear to
be at the center or at other locations due to projection effects. Much
of the chromospheric material either drains back to the surface or
heats to coronal temperatures or both during the early stages of the
eruption. This is likely the reason for the scarcity of in-situ
detections of such cool material in the inner heliosphere.

Finally, one may question the generality of our conclusions since we
have used as examples prominence eruptions from polar crown and
generally quiescent areas of the corona. This was done mainly for
convenience. Polar crown filament eruptions are both spectacular and
slow thus providing a large sample of clear highly detailed structures
at various stages of activity. It is much harder to find fast,
explosive events with clear structures due to their fast evolution and
the large disturbances they create in the surrounding corona. However,
detailed analyses of impulsive events do reach the same
conclusions. For example, \inlinecite{2010ApJ...724L.188P} were able to
detect the expanding cavity and follow the formation of the 3-part
structure within the center of an active region during a very
impulsive event (peak acceleration $\sim 1.5$ km s$^{-2}$). Similarly,
\inlinecite{2012ApJ...745L...5C} obtained clear distance-time plots for
an expanding cavity followed by a spectacular filament for an event
reaching (short-lived) accelerations close to 3 km s$^{-2}$.

\section{Revisting the LASCO Statistics: How Many FR-CMEs Are There?} \label{sec:stats}

Having addressed the origins of the various sub-structures of a CME in
the coronagraph and EUV images,we are now in position to answer one of
the most common questions on CME studies: How many FR-CMEs are there?
As noted in the introduction, the last studies to undertake that
question used \textsl{Solar Maximum Mission\/} (SMM) data
\cite{1993STIN...9326556B,2007ApJ...657..559K}. The answer ($\sim
$30\%) has been quoted ever since. But does it still hold after the
observations of thousands of CMEs? Besides, we want to make sure that
our examples of FR-indicative CME morphologies are indeed
representative of the CME phenomenon as a whole?

\subsection{Morphological Classfication of the CDAW Database}
\begin{figure}
\centerline{
\includegraphics[width=0.9\textwidth]{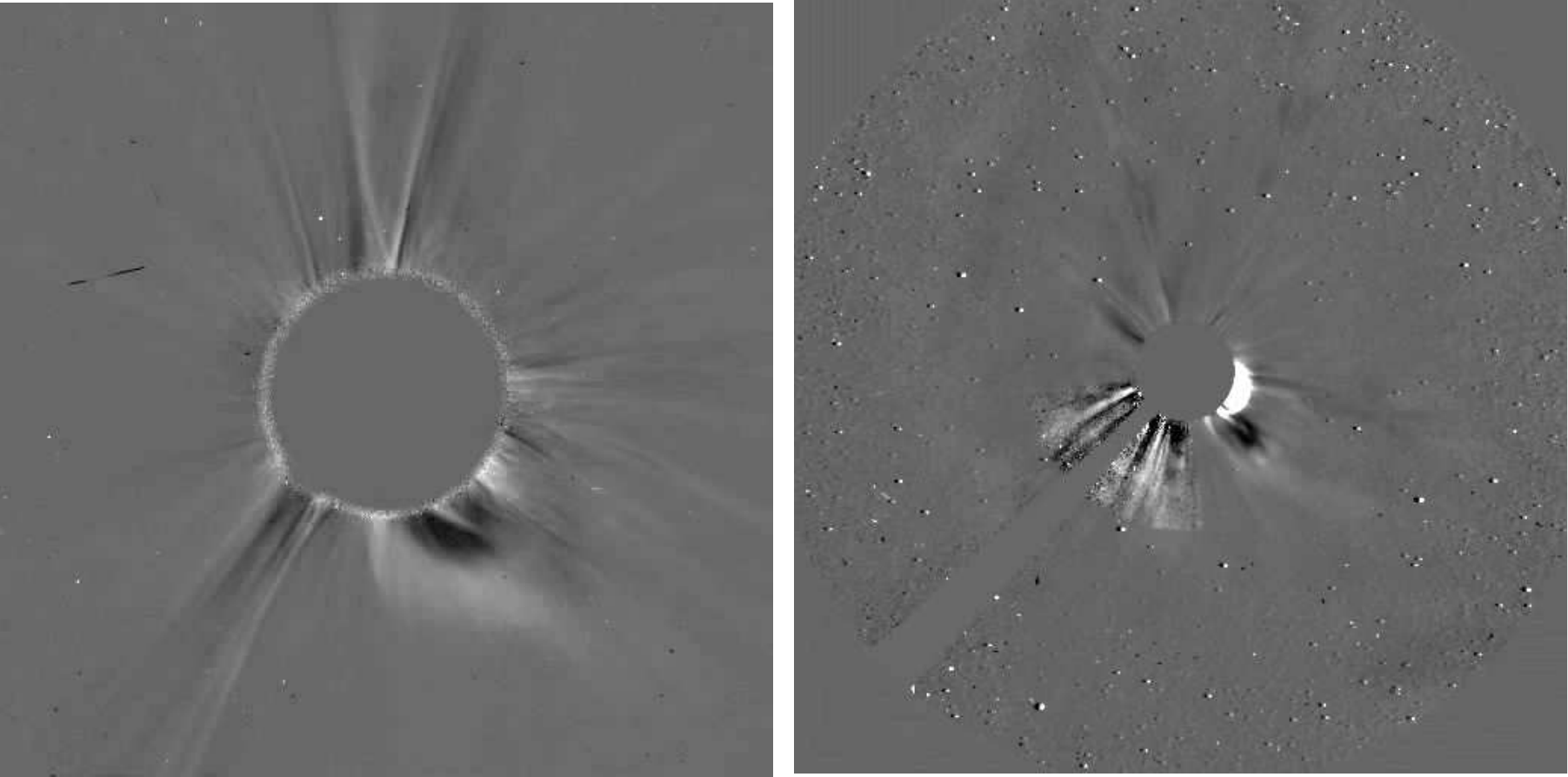}
}
\caption{\textsl{Left\/}: Example of an `Outflow' event. The CME lacks
  a 3-part morphology, it is too wide to be a `Jet' event, and the
  front is not sharp enough to be classified as a `Loop' CME.
  \textsl{Right\/}: Example of an `Unknown' category event. The event
  propagating along the C3 occulter does not show any of the
  characteristics of a CME and it is too faint to provide accurate
  measurements for any parameter. Such events should not be included in
  CME lists. Both images are snapshots from the online
  movies.  }\label{fig:stat_ex}
\end{figure}

So we have undertaken the task to go over the
full CDAW database and visually classify the events according to their
moprhology. Based on our discussion so far, and on our personal
experience with the LASCO images, we decided to classify the events
into five categories (compared to the ten categories in
\opencite{1985JGR....90.8173H}) as follows:
\begin{itemize}
\item\textsl{Flux Rope\/}: CMEs that exhibit a clear 3-part morphology
  (Figure~\ref{fig:3part}).
\item\textsl{Loop\/}: CMEs with a bright,filamentary loop but
  otherwise lacking a cavity and/or a core. Good indicators for the
  existence of shock (`two-front' morphology, Section~\ref{sec:shock})
\item\textsl{Jet\/}: Narrow CMEs ($\lesssim 40^\circ$) lacking a sharp
  front, detailed sub-structure, or circular morphology.
\textbf{\item\textsl{Failed\/}: Events that disappear in the C3 field of view despite being bright enough in the C2 field to be labelled as 'CMEs'. Their disappearance cannot be explained by lack of observations, overlapping CMEs, or other instrumental reasons. These events were discussed in \inlinecite{Vourlidas10}.}
\item\textsl{Outflow\/}: Events wider that jets, without clear loop
  front or cavity. They can as large as regular CMEs and can contain filamentary material (Figure~\ref{fig:stat_ex}, left).
\item\textsl{Unknown\/}: This `catch-all' category contains mostly
  wrongly identified events, events with too few observations ($< 4$),
  and events that cannot be classified in any of the other categories
  due to poor observations, such as presence of cosmic rays or data
  dropouts (Figure~\ref{fig:stat_ex}, right).
\end{itemize}

First, we excluded events with width less than $20^\circ$ and with
less than four observations. To do the classification, we used only
mass images \cite{Vourlidas10}. We displayed all available mass images
(LASCO C2 and C3) for each event with the same contrast
($\pm5\times10^{10}$ gr/pix). When the morphology was not apparent, we
used a movie of the mass images to get a better sense for the
development of the event as a function of time and
distance. We excluded the events in the `Unknown' category from the
statistics since we consider them as erroneous and/or unreliable
detections. Our results are shown in Table~\ref{tbl:stats}.

\subsection{Statistical Results}
For the purposes of this work, we analyzed six out of the sixteen
years of available CDAW mass measurements spread over Cycle 23. This
is a sufficiently large sample to extract robust estimates for the
rates of occurence of the various morphologies. We plan to report on
the statistics of the full CDAW catalog in a forthcoming publication.
 \begin{table}
   \caption{Time history of the CME eruption as marked by several key
     events.} \label{tbl:stats} 
 \begin{tabular}{lrrrrrrr}
Type      & 1997 & 1998 & 2000 & 2002 & 2004 & 2008 & 2010\\
\hline
Flux Rope &29    &42    &86    &54    &53    &35    &154 \\
Loop      &31    &43    &191   &100   &79    &9     &56\\
Outflow   &48    &96    &282   &145   &126   &56    &209\\
Jet       &5     &16    &85    &17    &32    &5     &32\\
Failed    &10    &33    &54    &10    &67    &14    &30 \\
Unknown   &74    &98    &276   &161   &220   &42    &113\\
\hline
Total     &224   &357   &984   &597   &587   &162   &476\\
FR-CMEs   &40\%  &33\%  &39\%  &35\%  &36\%  &37\%  &43\%\\
 \hline
 \end{tabular}
 \end{table}

One of the first interesting results from this exercise is the rather
large percentage of `Unknown' events. They comprise 984 out of 3387
events or 29\% of the sample. Recall that the `Unknown' category
includes events that are not CMEs or even outflows, events that are
part of a larger CME and should not count as separate entries, and
events with measurements ending prematurely. In other words, these
events should not be counted in the statistics or other analyses of
the CDAW catalog. 

Excluding the `Unknown' events, we find that the class of unambiguous
FR-CMEs, \textbf{\textsl{which consists of the sum of 'Flux Rope' plus 'Loop' classes\/}},
comprises 40\% of the total number of CMEs (962/2403 events). There is
no obvious correlation with solar cycle but there is a slight
hint. The highest percentages of FR-CMEs occur in 1997 and 2010. It
remains to be seen whether this result is statistically
significant. We emphasize that the 40\% of FR-CMEs is a
\textsl{lower\/} limit for the existence of FRs. Some of the `Outflow'
events could be FRs. Indeed many contain hints of `3-part' structures
or cavities, but they are either too faint or the background corona is
too disturbed by previous events, to make a conclusive
classification. So we choose to err on the side of caution and not
include them in the FR-related classes, at this point.

\section{Discussion and Conclusions} \label{sec:discussion}

Our aim is to provide convincing evidence of the CME as an erupting
FR. To that end, we have used a variety of EUV and white light
observations, MHD simulations, statistics, and have considered
projection effects and theoretical predictions. Leaving the question
of CME initiation aside, we found that the following picture can lead
to a self-consistent interpretation of the observations across many
wavelength ranges and is in agreement with the majority (if not all)
of our current theoretical understanding of explosive energy release
from the Sun.

Basically, a CME is the eruption of a magnetic flux rope with its
emission measure dominated by coronal temperature plasma, carrying a
prominence along its bottom dips, piling up the overlying streamer
plasma, and driving a wave ahead (if the acceleration is sufficently
high). This interpretation has long been adopted for the '3-part'-CMEs,
as we discussed earlier.

The novelty in this work is the interpretation of the bright loop
front as the pileup of material \textsl{at the boundary of the flux
  rope} irrespective of the `3-part' appearance. The interpretation is
supported strongly by the MHD simulations and straightforward physical
reasoning (Section~\ref{sec:models}). A FR structure propagating
through plasma presents an extended obstacle against which the
material is piled up and transported outwards. The narrow width and
brightness of that front further suggests that the pileup occurs over
a sharp boundary. Such a boundary is expected between the closed FR
fields and the ambient magnetic field. The sharpness of
the boundary may depend on the rate of magnetic field influx in the FR
during its formation or the initial acceleration and starting
height. Such effects have important connections to theories of CME
initiation and can be investigated now.

The other novelty is the introduction of the \textsl{`two-front'\/}
morphology by pointing out the existence of faint, relatively sharp,
fronts ahead of the bright loop front. The interpretation of the faint
front as density compression by a wave (or shock, depending on speed)
is again supported by MHD simulations, observations and physical
expectations. The stark observational differences between the bright
sharp front and diffuse front clearly point to a different origin. The
diffuse fronts are: well-defined, faint, followed by diffuse emission, can be
very extended, and envelope the sharp fronts. The sharp fronts, in
turn, are: sharp, bright, followed by emission depletions, have
well-defined extents, and are behind the diffuse fronts. The faint
fronts appear only during fast eruptions and their characteristics,
especially the weakness of their emission and lack of post-front
depletion are strong indications that these fronts are results of
local density compression and not of transported piled-up plasma. The,
albeit few, 3D reconstructions of the density profile across the front
\cite{2009ApJ...693..267O} can readily explaining the profile as a
result of LOS integration and recover compression ratios in agreement
with theoretical expectations (less than 4). Besides its importance
for understanding coronal shocks, the identification of the
`two-front' morphology allows an understanding of the geometry of halo CMEs
as it can help us distinguish among shock, streamer deflections and FR
signatures in the coronagraph images. In that way, we can now obtain
accurate outlines of the FR (or the shock, depending on the problem at
hand) which should lead to better inputs to CME propagation models.

The identification of these two features leads to a much simpler
classification of CME white light morphologies. We used four
categories (ignoring the `Unknown' category) compared to nine in
\inlinecite{1985JGR....90.8173H}. Two of them (`FR' and `Loop')
refer to the same FR instrinsic structure as we have argued. Jet-CMEs
also contain helical structures as recent research has shown
\cite{2008ApJ...680L..73P,2009ApJ...691...61P,2010AnGeo..28..687N}. Thus,
our classification is essentially reduced to events with and events
without \textsl{apparent} helical topologies. The helical topology may
not be visible in the latter for several reasons. They may propagate
at large angles from the POS \cite{2007ApJ...655.1142S} or through
areas disturbed by previous events. They may be too compact to discern
their cavity morphology without favorable projections
\cite{2006ApJ...650.1172W}. Finally some of these events do not appear
to be CMEs in the first place failing to reach large distances in the
corona (called `failed' CMEs by \opencite{Vourlidas10}). A certain
number of the remaining events appear to be related to H$\alpha$
and/or 304 \AA\ surges similar to the event studied by
\inlinecite{2003ApJ...598.1392V}. The low coronal signatures of these
events do not exhibit any particular morphology or geometry and hence
tend to appear as semi-amorphous clouds, with the occassional traces
of cool material.

Our final estimate of 41\% for the rate of occurrence of FR-CMEs in
the LASCO data may not look very differernt from the widely quoted
number of 30\%. However, one must first consider the size of the event
samples in past morphological works. \inlinecite{1979SoPh...61..201M}
reported a 26\% occurence of `Loop'-CMEs in a sample of 77 SMM CMEs
while \inlinecite{1984ARA&A..22..267W} found loop and bubble CMEs in
80\% of 65 SMM CMEs. Obviously, selection bias is important with such
small event samples. The largest morphological study to date
categorized 998 \textsl{Solwind\/} CMEs of which 31.3\% belonged to an
FR-CME class (we summed the statistics for the following structural
classes, curved front, loop, streamer blowout, fan)
\cite{1985JGR....90.8173H}. We base our statistics here on a sample of 2970
events, 3$\times$ larger than the \textsl{Solwind\/} sample and is still
expanding. We will classify the full LASCO database in the near
future. Therefore, we feel that our numbers in Table~1 are quite
robust and a large improvement over past work.

The central question of this \textsl{Topical Issue\/} is whether all
CMEs are flux ropes. To provide a conclusive answer (to the extent
possible in science), we attacked the problem in several ways: multiple
viewpoint coronagraphic observations of CMEs, multi-thermal EUV
observations of the pre-erupting structures, 3D MHD simulations, and
large sample statistics. We summarize our findings as follows:
\begin{itemize}
\item The detection of a bright filamentary front in CMEs is a clear
  indication of the existence of a FR even if the event does not
  exhibit the classical 3-part morphology.
\item At least $41\%$ of CMEs exhibit clear FR signatures (`3-part or
  `loop') in the coronagraph images. 
\item The `two-front' morphology (faint front followed by a bright
  loop) is a reliable indicator of a CME-driven wave (or shock,
  depending on speed).
\item The FR can be separated from the shock signatures in images of
  halo CMEs at least in locations where the bright loop appears.
\item  MHD simulations are able to capture the main structural
  properties of white light CMEs. 
\item The prominence is not the cavity and is not the FR but is the
  core. The cool prominence material rests on the dips of the field
  lines comprising the FR (in the case of pre-existing FR, at least).
\item The majority of the prominence material either drains to
  the surface or is heated to coronal temperatures during the early
  phases of the eruption. This may be the reason for the scarcity of
  in-situ detections of cool material.
\item A typical fast CME comprises five parts: shock front, diffuse
  sheath, bright front, cavity, and core. 
\end{itemize}

Our discussion suggests that it is time to rethink the original
definition for a CME \cite{1984JGR....89.2639H}, as expressed in
\opencite{2006LRSP....3....2S}: \textsl{``We define a CME to be an
  observable change in coronal structure that 1) occurs on a time
  scale of a few minutes and several hours and 2) involves the
  appearance (and outward motion) of a new, discrete, bright, white
  light feature in the coronagraph field of view.''} This definition
manages to be broad (no mention of the physical origin or nature of
the `structure') and narrow (CME is defined as a white light feature
observed by a coronagraph) at the same time. It may have been an appropriate
definition during the times of exploratory CME research, sparse
wavelength coverage, and simplified physical models. But times have
changed. We are regularly studying CMEs with multiple instruments and
wavelengths, have accumulated CME observations spanning a full solar
cycle, and are asking highly detailed questions with their
modeling. Thus, it may be useful to derive a more precise CME
definition using physically-based terms, at least for the events
exhibiting clear FR structures (FR-CMEs). Based on the work presented
here and in \inlinecite{Vourlidas10}, we propose the 
following definition:

\textsl{We define an FR-CME to be the eruption of a coherent magnetic,
  twist-carrying coronal structure with angular width of at least
  40$^\circ$ and able to reach beyond 10 
  R$_{\odot}$ which occurs on a time scale of a few minutes to several hours}. 

The next challenge now is whether we can apply this definition to all
CMEs (hence replace `FR-CME' with `CME' above). In other words, we
propose that the proper questions we should be asking is not
\textsl{`are all CME flux ropes?'\/} but rather \textsl{`Are there
  \underline{any\/} CMEs that are not FR-CMEs?'}

\begin{acks}
  The work of AV and RAH is supported by NASA contract S-136361-Y to
  the Naval Research Laboratory. BJL and YL acknowledge support from
  AFOSR YIP FA9550-11-1-0048, NASA NNX11AJ65G, and NNX08AJ04G. We
  thank G. Stenborg for his continuing efforts to provide better
  quality solar images. SOHO is an international collaboration between
  NASA and ESA. LASCO was constructed by a consortium of institutions:
  the Naval Research Laboratory (Washington, DC, USA), the
  Max-Planck-Institut fur Aeronomie (Katlenburg-Lindau, Germany), the
  Laboratoire d'Astronomie Spatiale (Marseille, France) and the
  University of Birmingham (Birmingham, UK). The LASCO CME catalog is
  generated and maintained at the CDAW Data Center by NASA and The
  Catholic University of America in cooperation with the Naval
  Research Laboratory. The SECCHI data are produced by an
  international consortium of the NRL, LMSAL and NASA GSFC (USA), RAL
  and Univ.  Bham (UK), MPS (Germany), CSL (Belgium), IOTA and IAS
  (France).

\end{acks}

\end{article}

\end{document}